\documentclass{aa}
\usepackage{natbib,psfig,graphicx,ulem}
\usepackage{txfonts}
\usepackage{soul,color}

\def\mathnew{\mathsurround=0pt}
\def\simov#1#2{\lower .5pt\vbox{\baselineskip0pt \lineskip-.5pt
\ialign{$\mathnew#1\hfil##\hfil$\crcr#2\crcr\sim\crcr}}}

\def\MeV{Me\kern-0.11em V}
\def\keV{ke\kern-0.11em V}

\begin{document}

\title{On the nature of faint Low Surface Brightness galaxies in the Coma
  cluster
\thanks{Based on observations obtained with MegaPrime/MegaCam, a joint project
  of CFHT and CEA/DAPNIA, at the Canada-France-Hawaii Telescope (CFHT) which
  is operated by the National Research Council (NRC) of Canada, the Institute
  National des Sciences de l'Univers of the Centre National de la Recherche
  Scientifique of France, and the University of Hawaii.}}

\offprints{C. Adami \email{christophe.adami@oamp.fr}}

\author{ C. Adami\inst{1} \and
R.~Pell\'o\inst{2} \and
M. P.  Ulmer\inst{1,3} \and
J.C. Cuillandre\inst{4} \and
F. Durret\inst{5} \and
A. Mazure\inst{1} \and
J.P. Picat\inst{2} \and
R. Scheidegger\inst{3} 
 }

\institute{
LAM, P\^ole de l'Etoile Site de Ch\^ateau-Gombert,
38, rue Fr\'ed\'eric Joliot-Curie,
13388 Marseille Cedex 13, France
\and
Laboratoire d'Astrophysique de Toulouse-Tarbes, Universit\'e de Toulouse,
CNRS, 14 Av. Edouard Belin, 
31400 Toulouse, France
\and
Department of Physics and Astronomy, Northwestern University,
2131 Sheridan Road, Evanston IL 60208-2900, USA
\and
Canada-France-Hawaii Telescope Corporation, Kamuela, HI 96743
\and
Institut d'Astrophysique de Paris, CNRS, UMR~7095, Universit\'e Pierre
et Marie Curie, 98bis Bd Arago, 75014 Paris, France
}

\date{Accepted . Received ; Draft printed: \today}

\authorrunning{Adami et al}

\titlerunning{On the nature of faint Low Surface Brightness galaxies in the Coma cluster}

\sethlcolor{red}

\abstract
{This project is the continuation of our study of faint Low Surface Brightness
  Galaxies (fLSBs) 
in one of the densest nearby ($z = 0.023$) galaxy regions known, the Coma
cluster.}
{Our goal is to improve our understanding of the nature of these objects 
by comparing the broad band spectral energy distribution with population
synthesis models, in order to infer ages, dust extinction
and spectral characteristics.} 
{The data were obtained with the MEGACAM and CFH12K cameras at the
CFHT. We used the resulting photometry in 5 broad band filters (u*, B, V, R, and I),
that included new u*-band data, to fit spectral models.  With these
spectral fits we inferred  a cluster membership criterium, as well
as the
ages, dust extinctions, and photometric types of these fLSBs.}
{ We show that about half of the Coma cluster fLSBs have a spectral
energy distribution well represented in our template library (Best Fit
fLSBs, BF) while
the other half present a flux deficit at ultraviolet wavelengths
(Moderately Good Fit
fLSBs, MGF). Among the BF fLSBs, $\sim$80$\%$ are probably part of the
Coma cluster based on their spectral energy distribution. BF fLSBs are
relatively young (younger than 2.3 Gyrs for 90$\%$ of the sample)
non-starburst objects. The later their type, the younger fLSBs are. A
significant part of the BF fLSBs are quite dusty objects (1/3 have A$_V$
greater than 1.5). BF fLSBs are low stellar mass objects (the later their
type the less massive they are), with stellar masses
comparable to globular clusters for the faintest ones. Their characteristics 
are partly correlated
with infall directions, confirming the disruptive origin for at least 
part of them.}
{}

\maketitle

\section{Introduction}\label{sec:intro}

 Faint Low Surface Brightness Galaxies (fLSBs) are a class of
galaxies residing both in rich clusters and in the field. This class of
galaxies is defined on the basis of its R band total magnitude (M$_R$$\geq$-14) and
its low surface brightness (R central surface brightness fainter than
$\sim$24, see Adami et al. 2006a for more details).
While they are not easily detectable, their presence can strongly constrain 
both hierarchical galaxy formation and
evolution and the amount of dark matter present in clusters
(e.g. Ulmer et al. 1996).

In terms of how fLSBs connect to the study of galaxy evolution in
clusters, Adami et al. (2006a) proposed the existence of three
different populations of fLSBs.  One population is analogous to the
red sequence of the color magnitude relation (CMR) found at much
brighter magnitudes by e.g. L\'{o}pez-Cruz et al. (2004). This is
probably the primordial fLSB population in the cluster. The second is
much redder and the third much bluer. Adami et al. (2006a) assumed
 that galaxies falling on the extended CMR had passively evolved and
were formed at the same time as most of the larger galaxies (excluding
recent infalls). In agreement with the standard model used to
explain the CMR (e.g. Kodama \& Arimoto, 1997), this means that these
CMR fLSBs underwent an initial burst of star formation that
consumed enough gas to stop any further star formation. The
metallicity of the CMR fLSBs is then expected to produce less red colors
than those of galaxies of the same age that could have retained their
gas better. In this picture, the fLSBs that fall on the red side of the
CMR would originate from giant stripped galaxies that were able to
retain their metals better before being disrupted. The fLSBs on the blue
side of the CMR probably come from recently infalling blue galaxies that are
just undergoing induced star formation, explaining the blue colors.
  
The primordial galaxies lying on the CMR would make up the low mass tail
of the galaxy mass distribution, as resulting from hierarchical galaxy
formation in CDM models (see for example, White \& Rees 1978; White \&
Frenk, 1991). In those models many low mass galaxies have never
coalesced into larger galaxies.  This scenario can be directly compared
to observations. The results have been a point of contention as to
whether there are enough observed dwarf galaxies compared to the
predictions of $\Lambda$CDM models or not (see for example, Davies et al
2004, Kravtsov et al. 2004, Roberts et al. 2004, 2007 and references
therein).  Hence, the study of low mass galaxies is directly linked to
cosmology, models of galaxy evolution, and large scale structure
formation.

In order to examine the hypotheses on the origin of the fLSBs found in
Adami et al. (2006a) we extend our initial study (based on deep B and R
data) to include the analysis of deep u*, V, and I images.  We require
that fLSBs be detectable in all bands.  With 5 bands we are able to fit
spectral distribution models from about 3500 \AA\ to 9000 \AA.  We then
compare our new results with the basic hypotheses that were presented in
Adami et al. (2006a) by fitting spectral energy distribution (SED)
models to our 5 colors.

Furthermore, Coma itself is special, not only from an observational
point of view since it is the closest dense cluster that is well out of
the galactic plane, but it contains one of the hottest, densest intra
cluster media and shows evidence for recent subcluster infall
(e.g. Neumann et al. 2003 or Adami et al. 2005a). We thus expect to
find evidence for the effects of these properties on the fLSB
population, which in turn gives us insight on the formation of
these fLSBs.

The redshift of Coma is 0.023, and for this paper we will assume a distance to
Coma of 100 Mpc, H$_0$ = 70 km s$^{-1}$ Mpc$^{-1}$, $\Omega _\Lambda =0.7$,
$\Omega _m =0.3$, distance modulus = 35.00, and therefore the scale is 0.46
kpc arcsec$^{-1}$. All magnitudes are given in the Vega system.

\section{Data and analysis method}

\begin{figure}[htb] \centering
\mbox{\psfig{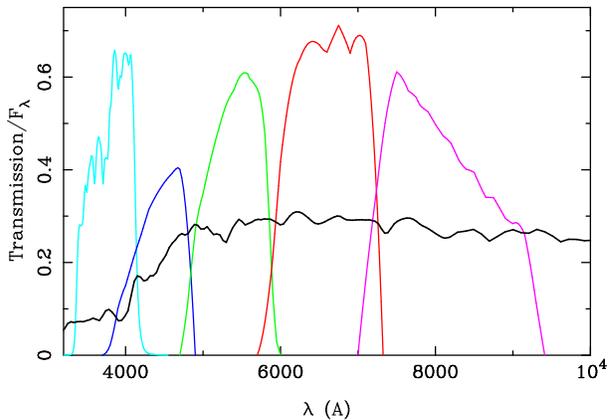}}
\caption[]{Filter + instrument transmissions for our set of filters (from left to
  right: u*, B, V, R, and I). The spectrum of an elliptical
  galaxy at z=0.023 is overplotted for comparison.}
\label{fig:filters}
\end{figure}

In the Adami et al. (2006a) work based on B and R band data, we used
statistical arguments and a comparison with blank fields to show
that $\sim$95$\%$ of the detected fLSBs were most probably members of
the Coma cluster. In the present paper we adopt a different approach to
check this result, using a photometric redshift technique, based on the
optimal wavelength coverage achieved by the u*BVRI data (see
Fig.~\ref{fig:filters}). The u* band filter is particularly useful
here because it allows to encompass the 4000\AA\ break, which is the
most important spectral feature in the considered redshift range for
both photometric redshifts and spectral classification of
galaxies. Therefore, instead of using statistical considerations
(e.g. Adami et al.\ 2006a), we derive the likelihood of being a cluster
member for each fLSB in this field.

\subsection{Data}

The data on which this paper is based are extensively described in Adami
et al. (2006b, B, V, R, and I bands) and in Adami et al. (2008, u* band,
hereafter A08). We only give here the salient points.

We observed the Coma cluster with the CFHT. First, we obtained a 2 field
mosaic with the CFH12K camera in 4 bands (B, V, R and I) in 1999 and
2000. These data cover a total field of 52$\times$42~arcmin$^2$ centered
on the two dominant cluster galaxies with a completeness level close to
R$\sim$24. These data are available at http://cencosw.oamp.fr.

Second, we obtained u$^*$ band data including the previous field in
2006 and 2007 with MegaPrime/MegaCam. The total exposure time was
9.66~hours, obtained by combining 58 individual spatially dithered
exposures of 10 min each.  The data reduction was performed with the
Terapix tools (http://terapix.iap.fr/), the standard procedure applied
for example to the CFHTLS fields (McCracken et al. 2008), and a
$\sim$1~deg$^2$ image was produced.

\subsection{Spectral template fitting method}

Photometric redshifts and derived parameters were computed for fLSB
galaxies using a modified version of the public code {\it
Hyperz}\footnote{http://webast.obs-mip.fr/hyperz/} (Bolzonella et al.\
2000; Ienna \& Pell\'o 2008), with the same settings as for the whole
Coma catalog (A08).  {\it Hyperz} is based on the
fitting of the photometric SED.  With respect to the public version, the
modified version\footnote{http:www.ast.obs-mip.fr/users/roser/hyperz}
includes a refined routine for flux integration which improves the
previous one in two different ways. The filter transmission curves are given
in photon units when needed, instead of the previous default value in
flux units (the user can actually combine both transmission definitions
within the same SED). The spectral resolution is better suited for
templates including strong narrow features. Several additional outputs
have been included to improve the cluster membership determination, such
as the integrated probability around the cluster redshift. Other
modifications mainly concern the inclusion of additional outputs
(e.g. the automatic classification of input rest-frame SEDs based on a
library of reference templates, the capability of deriving absolute
magnitudes in different filters based on two different scaling modes, as
well as the integrated probability around the best-fit photometric
redshift, and the normalized probability distribution in redshift), and
parameters (e.g. a galactic reddening correction for each object in the
input catalog instead of a global E(B-V), new templates and template
types, and the scaling of SEDs into physical units allowing the user to
easily compute stellar masses). This modified version is intended to
replace the previous public one.

 Photometric redshifts were computed in the range $0 \le z \le 6$, using a
large set of templates covering a broad domain in parameter space, mainly
defined by the spectral type, the age of the stellar population, and the
intrinsic reddening. The spectral library includes 14 templates for galaxies:

\begin{itemize}

\item Eight evolutionary synthetic SEDs computed with the last
version of the Bruzual \& Charlot code (Bruzual \& Charlot 1993, 2003),
      with 
the Chabrier
(2003) initial mass function (IMF) and solar metallicity: 
a delta burst [single-burst stellar population (SSP)], a constant star-forming
system, and 6 $\tau$-models with exponentially decaying star formation rate 
(SFR), roughly matching
the observed colors of local field galaxies from E to Im types. 

\item A set of 4 empirical SEDs compiled by Coleman et al.
(1980) to represent the local population
of galaxies, extended to wavelengths $\lambda \le
1400$\,\AA\ and $\lambda \ge 10000$\,\AA\ with the 
Bruzual \& Charlot spectra.

\item Two starburst galaxies (SB1 and SB2) from the Kinney et al. (1996)
template library.

\end{itemize}

SED fits were obtained with solar
metallicity templates. Low-resolution SEDs do not contain enough 
information to reliably fit all the parameters, given the degeneracies in 
the parameter space. Here we have 
fixed the IMF and metallicity, given their relatively smaller 
impact on the SED fitting results exploited in this paper as compared to 
the other relevant parameters, i.e. SFR, reddening and age of the stellar 
population, and their negligible impact on the photometric redshifts (see  
e.g. Bolzonella et al. 2000). Assuming smaller metallicities for the 
mean stellar population leads to older stellar ages and/or more extinction 
for the same SED fitting quality. Although the small strength of this 
effect does not change the observed behavior of this sample, one must 
bear in mind this limitation.   

  The internal reddening is considered as a free parameter with $A_V$
ranging between 0 and 2.0 magnitudes (E(B-V) between 0 and $\sim$ 0.6 mags),
according to Calzetti et al. (2000). We do not know if fLSBs are very dusty
objects and they could exhibit different properties compared to well studied
galaxies.

fLSBs in this sample are detected in all
the filters, and no prior was imposed in absolute magnitude. 

For each object we also obtained an estimate of its stellar mass based on the
best-fit of its SED at the redshift of Coma, using the Bruzual \& Charlot
templates mentioned above. A rough classification of the rest-frame
SED of galaxies into 5 different photometric types (spectral types
hereafter) was also obtained, based on the best fit with the simplest
empirical templates: (1) E/S0, (2) Sbc, (3) Scd, (4) Im and (5)
Starburst galaxies.

\subsection{Limitations}

There are four limitations of the described method: possible
uncertainties in the fLSB magnitude extraction, lack of adapted
templates to this peculiar class of galaxies, photometric redshift
probability distribution functions with complex shapes, and possible
degeneracies between output parameters.

\subsubsection{Magnitude uncertainties}

fLSBs are by definition very difficult objects to observe due to their
low surface brightness, usually close to the sky level. Simply using a
classical source extraction method (as the SExtractor package, Bertin
$\&$ Arnouts 1996) would probably lead to biased results due to the
difficulty to subtract the background level. In order to solve this
problem we applied a dedicated extraction code that already proved to be
efficient (see Ulmer et al. 1996 or Adami et al. 2006a for details) for
these purposes. In a few words, this IDL based package ingests a
SExtractor input catalog to select potential fLSB candidates and measure
their
magnitudes. This measure is done by fitting dedicated
surface brightness profiles (gaussians or exponentials depending on the
object characteristics and data seeing) and estimating the background in
hand-made external annuli. Each object requires manual inspection,
thus allowing for a given object to deal with peculiar problems such as
close neighbors or CCD defects.

This process was followed homogeneously in the five bands (u*, B, V, R,
and I) with the same settings as in Adami et al. (2006a) and this
ensures the best possible estimates for the fLSBs magnitudes. These
magnitudes were fed into the photometric redshift code.

\subsubsection{Templates}

\begin{figure*}[hbt]
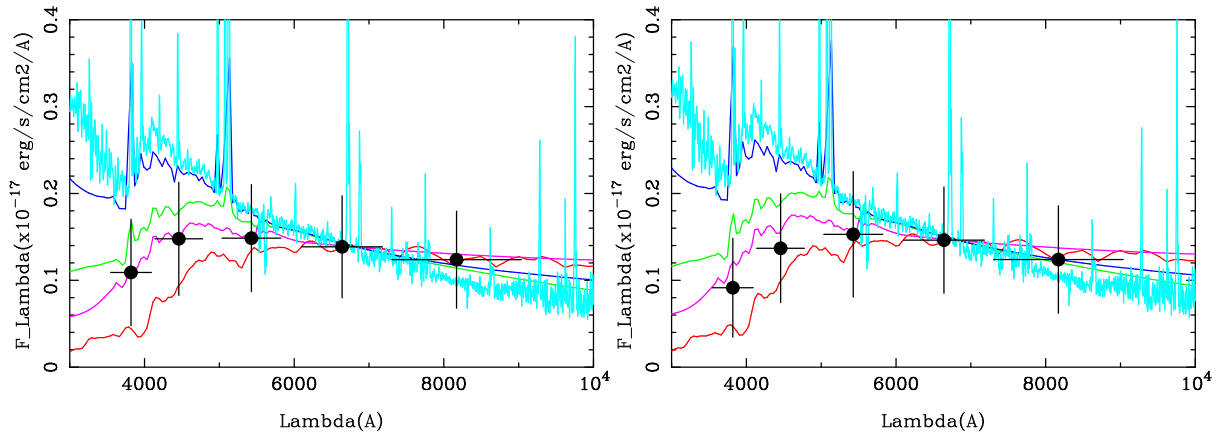
 
\centering
\mbox{\psfig{figure=LSB_good_sed.ps,width=8.cm,angle=270}\psfig{figure=LSB_kk_sed.ps,width=8.cm,angle=270}}
\caption[]{Mean spectral energy distribution in u$^*$BVRI for the different
  fLSBs (right panel: MGF fLSBs, left panel: BF fLSBs), overplotted on the
different SEDs: red, pink, green, dark blue and
cyan correspond to types 1 (early types) to 5 (star burst galaxies) 
respectively. SEDs for the BF and MGF populations are displayed as black dots. 
Photometric and template SEDs are arbitrarily normalized in the
R band flux for BF fLSBs. Note that error bars on fluxes correspond to 1$\sigma$
rms values with 
respect to the mean colors instead of photometric
errors. These plots include all the detected fLSBs, disregarding their
$Hyperz$ estimated probability to be Coma members.}
\label{fig:figSED}
\end{figure*}

In view of their faintness (R$\geq$21) fLSBs are up to now basically
spectroscopically unknown objects. The template library may therefore be 
incomplete.  It is beyond the scope of this paper to
determine the SEDs of the whole fLSB population (spectroscopically or with
extensive narrow-band photometry), so we chose to study
only the fLSBs that were reasonably represented in our template
library.

This was made in two steps. First, we assumed the Adami et al. (2006a)
results showing that nearly all fLSBs along the Coma cluster line of
sight were likely to be cluster members. We therefore assigned the Coma
cluster redshift to all fLSBs. Then, we compared the observed magnitudes
with our template library, computing the best fit reduced $\chi ^2$ and
probability P($\chi ^2$). This analysis shows that part of our fLSBs exhibit a
flux deficit in the u* band compared to the mean spectral type of these
objects (Sbc or type 2).

This led us to split our sample in two parts: the best fit sample
(hereafter BF: not showing the UV deficit, see Fig.~\ref{fig:figSED})
and the moderately good fit sample (hereafter MGF: showing this deficit,
see also Fig.~\ref{fig:figSED}) of the synthetic templates to the
observed SEDs (on the basis of $\chi ^2$ and P($\chi ^2$) being greater
or lower than 1.7 and 10$\%$ respectively). Note that error bars in the
SED displayed in Fig.~\ref{fig:figSED} correspond to 1$\sigma$ rms
values with respect to the mean fluxes of the fLSB populations.

Regarding the MGF fLSBs, we will study these objects without using the template
fit parameters (as spectral type, age, or extinction). This class includes by
definition the small percentage of fLSBs (predicted in Adami et al. 2006a) 
that are not Coma members. We removed all these
galaxies from the master sample (MS in the following). The MS will therefore
contain only BF fLSBs (312 galaxies).

\subsubsection{Redshift probability distribution function}

\begin{figure}[hbt] 
\centering \mbox{\psfig{figure=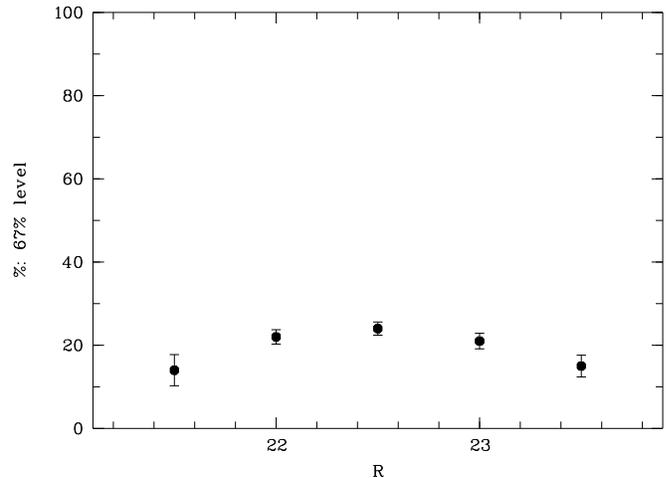,width=9.5cm,angle=270}}
\caption[]{Percentage of the best fit (BF) sample that can be excluded
from being in the $0 \le z \le 0.12$ interval, based on the quality of
the SED fit at the 67\% confidence level as a function of the galaxy
magnitude.}  \label{fig:quality}
\end{figure}

In order to test the fLSB Coma membership, we could have considered the
direct photometric redshift estimate. The quality of photometric
redshifts achieved for the present photometric catalog has been studied
in detail in A08, using both {\it Hyperz} and {\it LePhare} (e.g. Ilbert
et al. 2006). However, as also discussed in A08, direct photometric
redshift estimates are subject to several biases, in particular when the
photometric redshift probability distribution function presents several
peaks along the redshift interval. We therefore prefered to consider (as
in A08) the redshift Probability Distribution Function (PDF: a function
giving the probability for a given galaxy to be at a given
redshift). This allowed us to compute the integrated probability for
each galaxy of being at the cluster redshift $\pm0.1$ (actually within
the $0 \le z \le 0.12$ interval), based on their normalized probability
distributions within the $0 \le z \le 2$ interval.

We cannot exclude galaxies in the $0 \le z \le 0.12$ interval and not
belonging to the Coma cluster. However, if we consider the luminosity
functions of Ilbert et al. (2005) between R=21 and 24 (the approximate
magnitude interval in this paper) and assume that fLSBs follow the
normal galaxy luminosity function, we can show that the number of field
galaxies in the cosmological comoving volume included in our field of
view and at redshift lower than 0.12 is $\sim$15 times lower than the
number of Coma cluster galaxies. Moreover, we have shown in Adami et
al. (2006a) that field fLSBs are much rarer than in the Coma
cluster. This implies that the ratio between the number of Coma cluster
fLSBs and $0 \le z \le 0.12$ fLSBs is probably even larger than 15. We
therefore estimate than less than 5$\%$ of the fLSBs classified as Coma
members are in fact non cluster members at z$\leq$0.12.

Fig.~\ref{fig:quality} gives the percentage of the BF sample that can be
excluded from being in the $0 \le z \le 0.12$ interval at the 67\% confidence level (we
exclude a given fLSB from the sample if the probability that it is in
the $0 \le z \le 0.12$ interval is smaller than 67$\%$) as a
function of the galaxy magnitude. Even if this result is consistent with
a large fraction of the BF fLSBs being genuine Coma members, as
previously demonstrated in Adami et al. (2006a), the galaxies with a
membership probability lower than 67\% will be removed from the MS.

At this point, the MS includes 256 BF fLSBs which are very likely Coma members.

\subsubsection{Degeneracies between output parameters}

\begin{figure*}[htb] \centering
\mbox{\psfig{figure=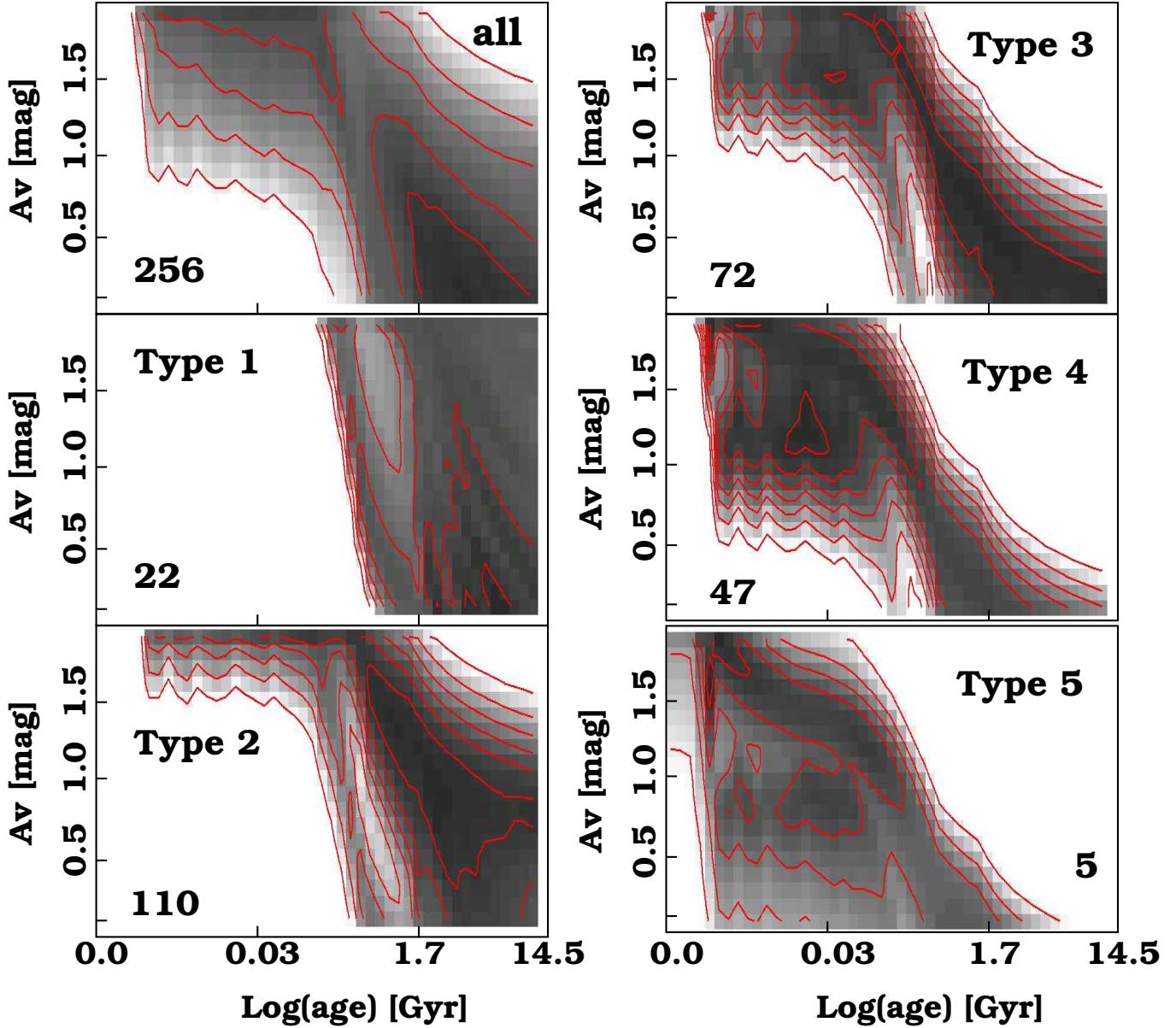,width=18.cm,angle=0}}
\caption[]{This figure displays for the master sample (MS) the averaged
likelihood maps and degeneracy in the (age, A$_V$) parameter space for
the different best-fit spectral types. A logarithmic scale is used for
ages, while A$_V$ is given in magnitudes. Each (age, A$_V$) map is the
average of individual maps for all fLSBs in a given class, where each
individual map was arbitrarily normalized to 1 at the minimum $\chi2$
value before maps were combined (see text for more details). The numbers
at the bottom of each panel correspond to the total number of fLSBs
combined. The most probable regions are shown in dark, on a logarithmic
scale. Solid lines display increasing isodensity contours in the (age,
A$_V$) map in a linear scale ranging between 0 and 1, with 0.1 steps.}
\label{fig:age_Av}
\end{figure*}

Our experience shows that the only significant degeneracy can occur between
the age and A$_V$. Starting from the MS (fLSBs likely to be at the Coma
redshift and well represented in our template library), we have then computed
likelihood maps in the parameter space defined by galaxy type, age and
reddening.  In particular, likelihood maps in the (age, A$_V$) parameter
plane were obtained for each galaxy and for each evolutionary Bruzual \&
Charlot template given above. 
 
The fitting parameter ages and values of A$_V$ are partly degenerate in
the sense that fits of comparable qualities can be obtained for
different sets of parameters. However, a large fraction of the Coma
fLSBs is best fit with reddening values above A$_V \sim 0.2$. To
investigate these degeneracies further, we have stacked together all the
combined (age, A$_V$) likelihood maps, for all and for different fLSB
populations according to their best-fit spectral types.  Results are
shown in Fig~\ref{fig:age_Av}. Each (age, A$_V$) map in this figure is
an averaged combination of individual maps for all fLSBs in a given
class. Before combining the maps, each individual map was normalized to
a value of 1 at the minimum $\chi2$ value. Maps displaying a low
contrast indicate a high dispersion in the individual maps, whereas maps
with a high contrast correspond to the regions where the galaxy
populations tend to concentrate.

Although template fitting can formally reach ages as old as 17.5 Gyr, 
in the following we limit the age of the stellar
population to the age of the universe. This assumption has no
consequence on the final conclusions.

\section{General results}

In this section, we first briefly describe the MGF fLSBs and we mainly
concentrate on BF fLSBs, the only ones for which spectral fitting results can
really be trusted.

\subsection{Moderately Good Fit (MGF) fLSBs}

The MGF fLSBs represent 59$\%$ of the initial sample. The spectral
shape of these objects is typical of galaxies  having experienced their
  last star burst several Gyrs in the past as shown by the UV flux deficit. This is in good
agreement with Adami et al. (2006a) who suggested that 2/3 of the total sample
could be galaxies formed in low mass dark matter halos. These objects would
have experienced a short burst of star formation, rapidly halted by
their inability to retain their gas against supernova winds. These
objects would then have passively evolved after the burst. 
The spatial distribution of the MGF fLSBs is not significantly
different (at the 99$\%$ level) from the complete sample of fLSBs, from a
Kolmogorov-Smirnov 2D test. There is no visible spatial concentration. This
is not surprising, as this population is probably relatively old (probably
part of the primordial fLSBs of the cluster) and had a long time to
homogenize. Even if MGF fLSBs do not allow accurate spectral parameters determination,
we can still quote that $HyperZ$ predicts a mean age of 2.2$\pm$3.7 Gyrs for this
population. This is of the order of the age of the oldest BF fLSB (see below).

We will now concentrate on the BF fLSBs.

\subsection{Best Fit (BF) fLSBs compared to the Adami et al. (2006a) classes}

\begin{figure*} 
\centering
\mbox{\psfig{figure=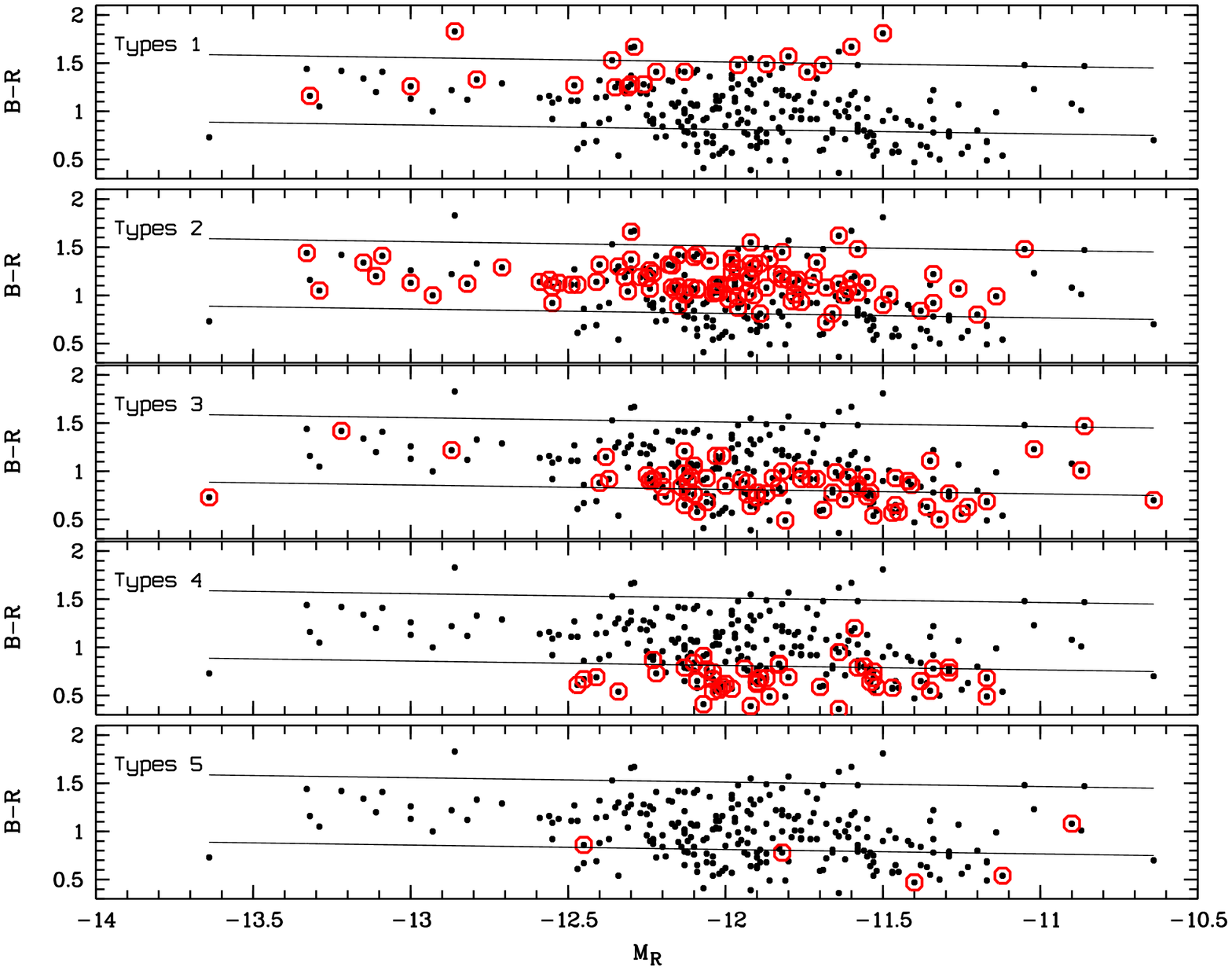,width=18.cm,angle=270}}
\caption[]{Master sample (MS) fLSB color magnitude relation as a
function of spectral type. The two inclined lines show the limits of the
red sequence defined in Adami et al. (2006a). Black dots are all the MS
fLSBs. Red circles are MS fLSBs with a given type (from top to bottom: types 1
to 5).}  \label{fig:CMRBF}
\end{figure*}

We defined in Adami et al. (2006a) three classes of fLSBs (bluer than
the Coma red sequence, on the Coma red sequence, and redder than the
Coma red sequence) that we tentatively identified with debris from
disrupted disk galaxies, with old and passively evolving primordial
cluster fLSBs, and with depleted cores of early type galaxies
respectively. In order to give the relation between these three types
and the five spectral types defined in the present paper for MS, we show
in Fig.~\ref{fig:CMRBF} the location of the different spectral types in
the color magnitude relation defined in Adami et al. (2006a). We see
that types 4 and 5 are mainly part of the blue fLSB class of Adami et
al. (2006a). Types 3 are distributed between the blue and the red
sequence fLSBs of Adami et al. (2006a). Types 2 are essentially red
sequence fLSBs from Adami et al. (2006a). Types 1 are distributed
between the red and the red-sequence fLSBs of Adami et al. (2006a).

\subsection{BF fLSB spectral characteristics}

     We see in Fig.~\ref{fig:age_Av} that type 1 fLSBs are older
than 0.5 Gyrs, whatever A$_V$. Type 2 fLSBs are also older than 0.5 Gyrs,
except when considering strong A$_V$. Type 3 fLSBs have various ages,
but are preferentially relatively old objects except, again, when considering
strong A$_V$. Type 4 fLSBs
also have various ages but are preferentially younger than 0.5
Gyrs. Finally, type 5 fLSBs are clearly young objects (in good agreement
with being starburst galaxies) whatever A$_V$. These results are
consistent with the general idea of galaxies being younger when the
type increases. Comparing our Fig.~\ref{fig:figSED}
with Fig.~1 of Asari et al. (2007) also shows that type 1 fLSBs are
probably high metallicity objects while late type fLSBs have lower
metallicities. However, we will not push this comparison further as we
will show that we are dealing here with significantly lower stellar
masses than Asari et al. (2007).

\begin{figure}[hbt] 
\centering \mbox{\psfig{figure=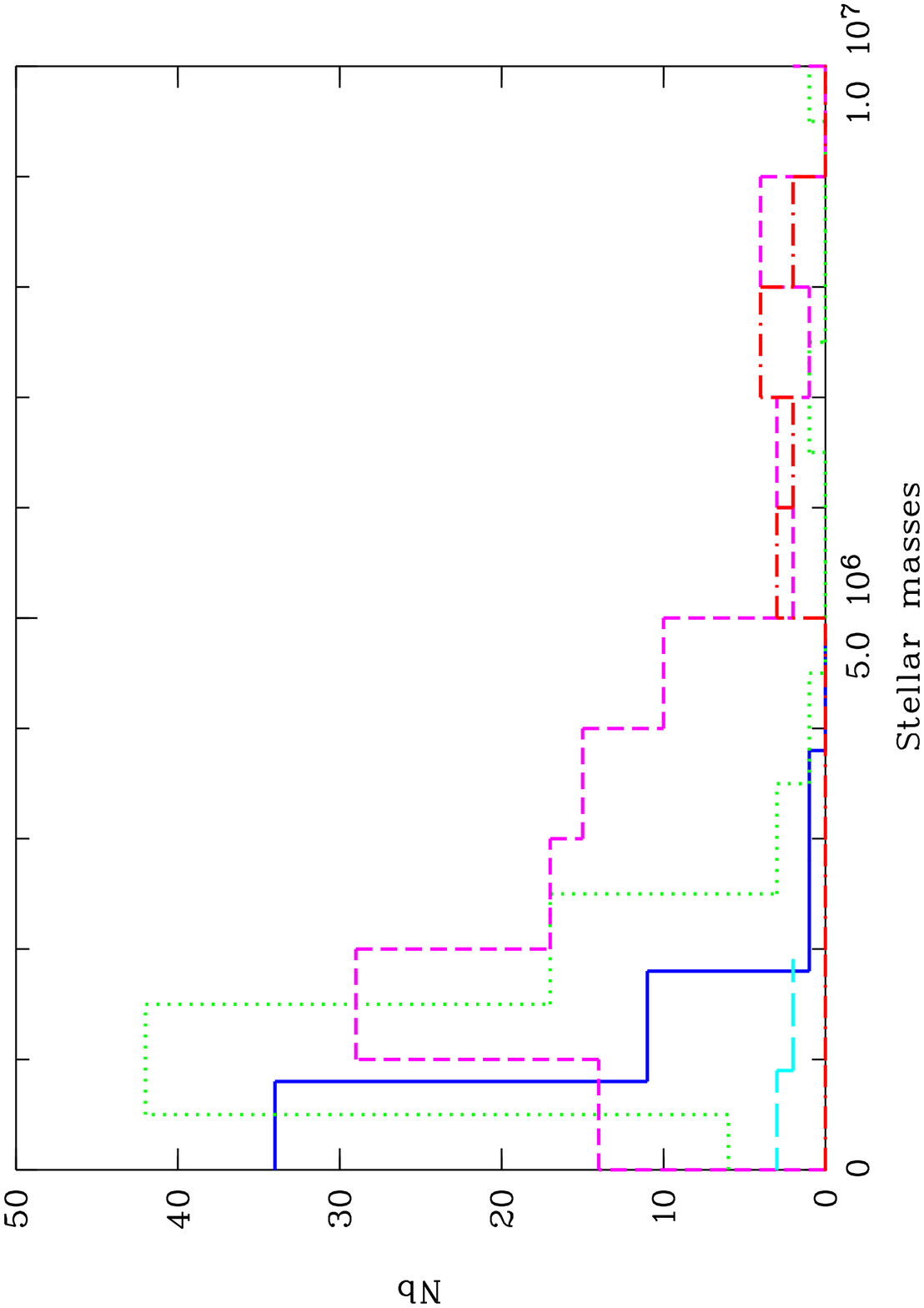,width=9.cm,angle=270}}
\caption[]{Histograms of the MS stellar masses (in M$_{\sun}$) for
different spectral types (type 5: cyan long dashed; type 4: blue
continuous; type 3: green dotted; type 2: pink short dashed; type 1:
red dot-dashed). } \label{fig:figstellar}
\end{figure}

\begin{figure}[hbt] 
\centering \mbox{\psfig{figure=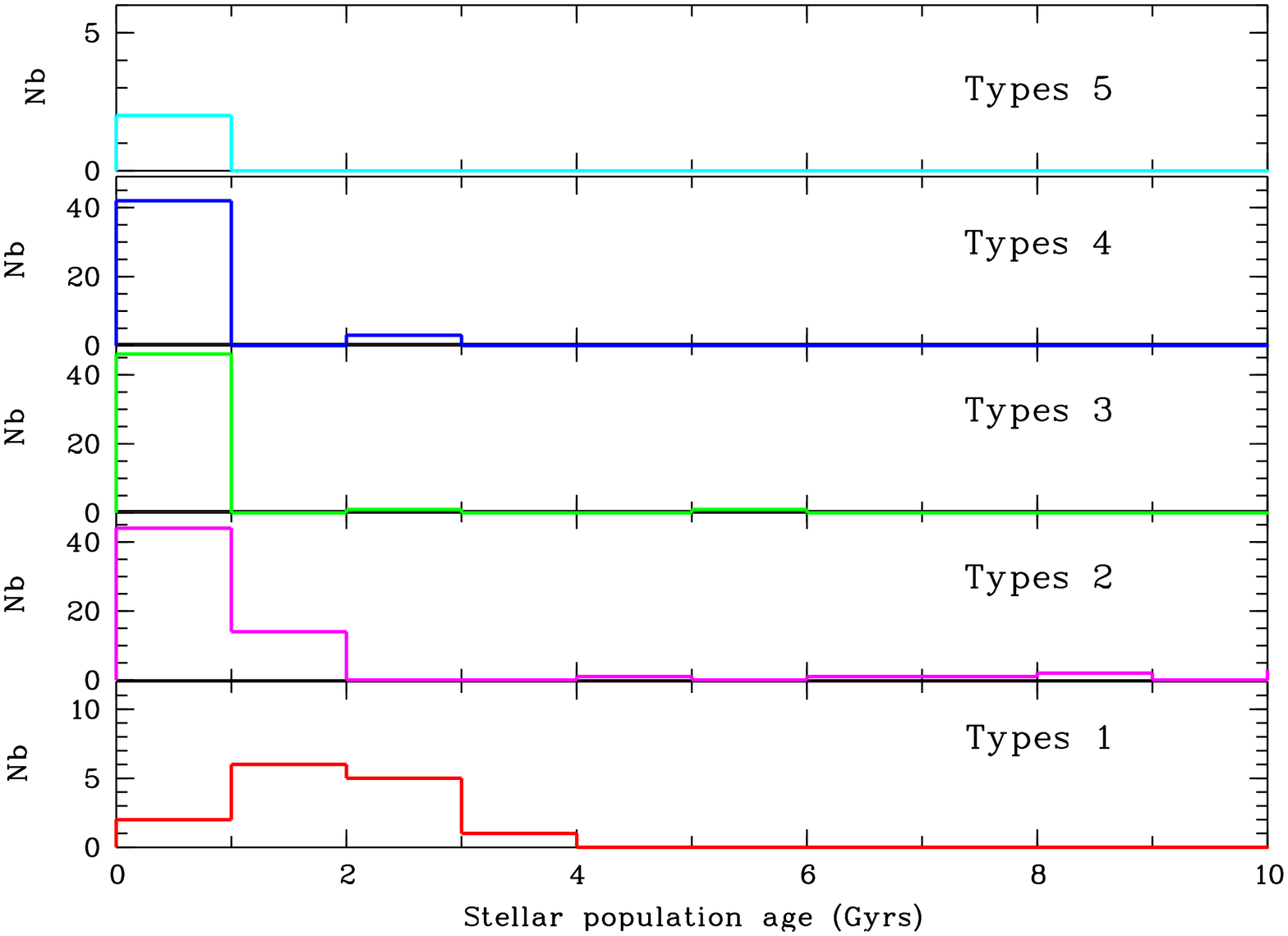,width=9.5cm,angle=270}}
\caption[]{Histograms of the MS ages for
different spectral types. From top to bottom, type 5: cyan, type 4: blue, 
type 3: green, type 2: pink, type 1: red. } \label{fig:figage}
\end{figure}

                The mean extinction is moderate and equal to
1.02$\pm$0.60. However, this includes quite dusty objects: 32$\%$ have
Av greater than 1.5.

fLSBs in this sample exhibit absolute magnitudes ranging between M$_B
\sim -9$ and $-12$. Since absolute magnitudes are directly related to
stellar mass, we show in Fig.~\ref{fig:figstellar} the histograms of the
MS fLSB stellar masses. First we confirm that we deal with a very low
stellar mass galaxy population, comparable to globular clusters for a
large part of the sample.
We also see that the earlier the type, the more massive the fLSBs in terms of
stellar mass.

Very few galaxies among the fLSB population display SEDs consistent with
starburst galaxies. Therefore, most of them are not undergoing an active
process of star formation. On the other hand, the MS fLSBs contain on
average quite a young stellar population (see Figs.~\ref{fig:age_Av} and
\ref{fig:figage}), the 90$\%$ youngest galaxies of the sample being
younger than 2.3 Gyrs; they are even found to be younger than 1.45 Gyrs
when excluding types 1, and younger than 0.7 Gyr when excluding types 1
and 2. Therefore these populations formed recently in the cluster
history.

The spatial distribution of the five starburst BF fLSBs 
does not show any particular pattern or correlation on large
scales. However, there is a strong correlation between these types and
the presence of close neighbors. Indeed, all these starburst BF fLSBs  have a close
neighbor, suggesting that these starbursts are induced by galaxy-galaxy
interactions. Spectroscopy will, however, be needed to go further in this study,
in particular to determine whether these close neighbors are physically
linked. This task is difficult because fLSBs are faint and diffuse
objects, and therefore this discussion is out of the scope of the present
paper.

\section{MS fLSB spatial distribution}

In the next three subsections, we investigate the spatial
distribution of BF fLSBs (MS) as a function of parameter
values (absolute magnitude, age, and spectral type). In order to achieve
this goal we computed the mean value of a given parameter as a
function of location in the field. This was done with an adaptative
kernel technique (e.g. Adami et al. 1998) that produces a
map of the mean parameter values, discretizing the field of view in several
pixels. This technique is a good balance between the spatial resolution 
and the uncertainty on the mean parameter value at a given place.

\subsection{Spatial distribution of MS fLSBs as a function of parameter
  values: Absolute magnitude} 

\begin{figure} 
\centering
\mbox{\psfig{figure=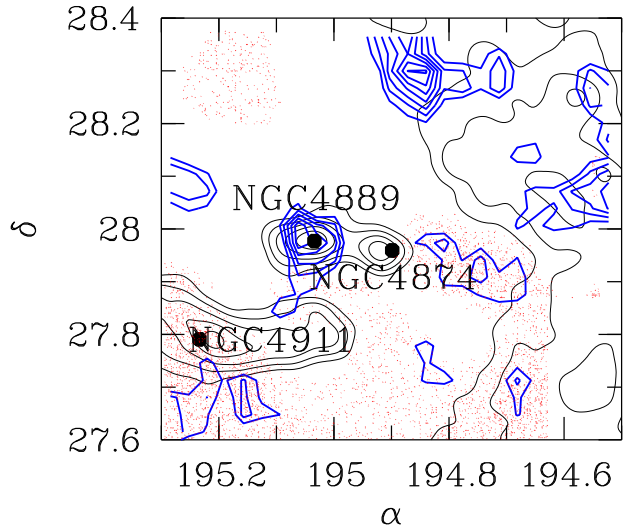,width=14.cm,angle=270}}
\caption[]{Red shaded areas show the regions where BF fLSBs with mean R
  absolute magnitudes between $-12$ and $-9.5$ are detected. The darker the
  shading, the brighter the mean magnitude in the [-12;-9.5] range. X-ray
  substructures (thin black contours) and significant fLSB
  concentrations from Adami et al. (2006a: thick blue contours) are
  overplotted.  Individual positions of fLSBs are not displayed. The lower
  right corner of the figure ($\alpha$$\leq$$\sim$194.55) is empty
  as we do not have B and V data in this region (see Adami et al. 2006b).}
\label{fig:figmag}
\end{figure}

The faintest fLSBs (Fig.~\ref{fig:figmag}, R$\geq$-12) are mainly distributed in the
south of the cluster. In the northern part of the cluster, the mean R
magnitude per pixel is almost always brigther than -12. This could be
put in perspective with the fact that this is also the less populated
region of the cluster in terms of faint galaxies
(not only fLSBs: see Adami et al. 2007). We therefore would like to know
if the masses of the potential MS fLSB progenitor galaxies have an
influence on the MS fLSB magnitudes. More violent interactions are
likely to take place in the south part of the cluster (because of the
NGC~4839 and NGC~4911 infalls) and could also produce smaller and
fainter galaxies. For example, Yoshida et al. (2008,
based on Subaru imaging data) have shown an example of such a violent
interaction in the south of the Coma cluster. This interaction produced knots
as faint as M$_R=-12$. However, such a general scenario still clearly
needs to be assessed for example by numerical simulations, even if this
is a demanding task (it requires to detect total masses of the order of
10$^7$~M$_\odot$  in simulations, see Adami et al. 2007).

The detection of fainter objects in the south (compared to the
north) of the cluster could also be due to instrumental effects
since our southern exposures are deeper in some bands, e.g. in the
R band (Adami et al. 2005b). However this is unlikely to explain
completely the previous effect, as we show in the same paper that
LSB detection levels are very similar in the north and south
regions.

\subsection{Age}

In section 2.3.4. we showed the degeneracy between age and extinction.  Here
we assume that the internal extinction of these galaxies is not
atypically high and that the age of the stellar population is the key
physical reason for the spectral differences among our sample.  The
oldest objects (Fig.~\ref{fig:figage}) are located to the
west of NGC~4874, between NGC~4911 and NGC~4889 and to the north-east of
NGC~4889.  NGC~4874 is probably the oldest dominant galaxy in the
cluster (e.g. Adami et al. 2005a). Therefore, we expect to have a relatively 
old population in its vicinity and this is, at least partially, the
case. However, ages between 0.8 and 3 Gyrs are not that old 
and this speaks in favor of a relatively recent
formation of these galaxies, with destruction of infalling giant galaxies as a possible
origin. The same explanation is also possible for the region between
NGC~4911 and NGC~4889 because this area is close to the cluster center
and probably subject to strong tidal forces that could be able to destroy
relatively bright galaxies. 

\begin{figure} 
\centering
\mbox{\psfig{figure=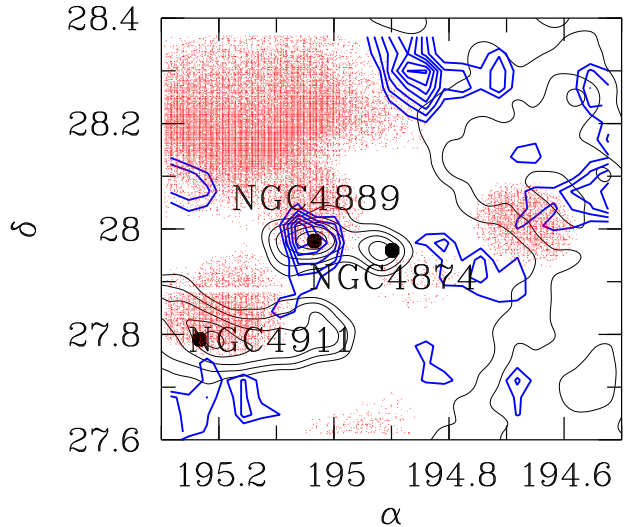,width=14.cm,angle=270}}
\caption[]{Red shaded areas show the areas where BF fLSBs with mean 
ages between  0.8 and 3 Gyrs are detected. The darker the
  shading, the older the mean age in the [0.8;3] Gyrs range. X-ray 
substructures (thin black contours) and 
significant fLSB
  concentrations from Adami et al. (2006a: thick blue contours) are
  overplotted. Individual positions of fLSBs are not displayed.}
\label{fig:figage}
\end{figure}

\subsection{Spectral type}

\begin{figure} 
\centering
\mbox{\psfig{figure=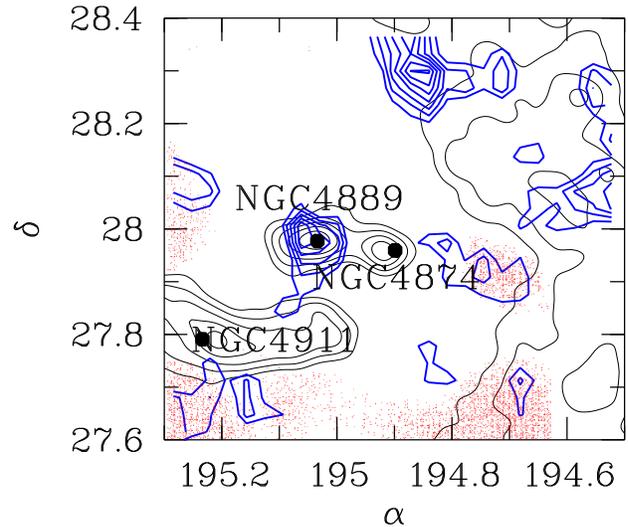,width=14.cm,angle=270}}
\caption[]{Red shaded areas show the regions where BF fLSBs with mean 
spectral types between 2.85 and 4 
are detected.  The darker the shading, the later the mean type in the 
[2.85;4] range.X-ray substructures (thin black contours) and significant fLSB
  concentrations from Adami et al. (2006a: thick blue contours) are
  overplotted. Individual positions of fLSBs are not displayed.}
\label{fig:figAvmod}
\end{figure}

The latest type BF fLSBs (Fig.~\ref{fig:figAvmod}) are mainly located in
front of the infalling layers at the west of the cluster, and at the
extreme south-east close to NGC~4911.  Finding the latest types in the
west infalling layers and along the NGC~4911 motion axis is in good
agreement with the origin suggested for the blue fLSBs detected in Adami
et al. (2006a): debris from spiral galaxies coming from the field.

\subsection{Spatial smoothing of the fLSBs}

We show in Figs.~\ref{fig:figmag}, ~\ref{fig:figage}, and
\ref{fig:figAvmod} that fLSB properties are most of the time correlated
with the main characteristic directions of the cluster (NGC~4911,
NGC~4839, western infalling layers and north-east direction). However,
other areas also exhibit peculiar fLSB populations in terms of A$_V$,
magnitude, age, or spectral type, without being located on any infall
directions. This is only possible through a rapid spatial smoothing of
the fLSBs. Populating the whole south part of the cluster with the
faintest fLSBs (Fig.~\ref{fig:figmag}) requires for example a spatial
migration of $\sim$650 kpc. The fLSB ages being mostly lower than 2.3
Gyrs, such a migration is only possible if fLSBs have a velocity
greater than 350 km/s. fLSBs can be primordial objects in the cluster or
disruptive galaxies (produced by larger galaxy disruptions).

Primordial fLSBs have velocity dispersions typical of galaxies with
similar masses. Adami et al. (1998) have shown that Coma cluster
galaxies fainter than M$_R=-17$ have a velocity dispersion of the order
of 1200 km/s. This is clearly sufficient to explain a $\sim$650 kpc
migration in 2.3 Gyrs.

Regarding disruptive fLSBs, their initial velocity has to be greater
than their parent galaxy escape velocity (otherwise fLSBs originating
from a major galaxy disruption will never escape) and the escape
velocity then has to be greater than 350 km/s.  This puts a constrain
on the mass of these fLSBs progenitors. Wu et al. (2008) show for
example that the escape velocity at the periphery of the Milky Way 
(1.9 10$^{12}$ M$_{\odot}$ in their paper) is $\sim$350 km/s. Less
massive galaxies (e.g. NGC 4861 or NGC 2366, of the order of 2 10$^{11}$
M$_{\odot}$, see van Eymeren et al. 2007) have a lower escape
velocity, lower than 50 km/s. This rough calculation therefore shows
that disruptive fLSBs progenitors are probably more massive than a few
10$^{12}$ M$_{\odot}$.

\section{Summary and conclusions}

We show in this paper that about half of the Coma cluster fLSBs have a spectral
energy distribution well represented in our template library (BF fLSBs) while
the other half present a flux deficit in the ultraviolet wavelengths (MGF
fLSBs).

MGF fLSBs are probably galaxies having experienced their last burst of star
formation several Gyrs ago. They are homogeneously distributed in the cluster
and are probably part of the Coma primordial fLSBs. 

Among the BF fLSBs, $\sim$80$\%$ of these objects are probably part of the
Coma cluster based on their spectral energy distribution. Their type closely
links them with the class defined in Adami et al (2006a). BF fLSBs are
relatively young (younger than 2.3 Gyrs for 90$\%$ of the sample)
non-starburst objects. The later their type, the younger fLSBs are. A
significant part of BF fLSBs are quite dusty objects (1/3 have A$_V$
greater than 1.5). They are low stellar mass objects (the later their
type the less massive they are), with stellar masses
comparable to globular clusters for the faintest fLSBs. Their 
characteristics are partly correlated with infall directions, 
confirming the disruptive origin for at least part of them.

We therefore confirm and refine a large part of the Adami et al. (2006a) 
conclusions. The
next step will be to get spectroscopy of the fLSBs. This will give us a direct
measure of the possible star-forming activity of these objects and allow us to draw a
more complete picture of the fLSB behavior in the Coma cluster.

\begin{acknowledgements}
The authors thank the referees for their time, and for useful and constructive
comments.
The authors are grateful to the CFHT and Terapix (for the use of QFITS, SCAMP
and SWARP) teams, and to
the French CNRS/PNG for financial support. Some of the authors
also acknowledge support from NASA Illinois space grant
NGT5-40073, from
Northwestern University and from NSF grant AST-0205960.\\
\end{acknowledgements}

\end{document}